\documentclass[superscriptaddress,aps,prb,twocolumn]{revtex4-2}

\usepackage{graphicx}

\usepackage{soul}
\usepackage{amsmath}
\usepackage{bm}
\usepackage{siunitx}
\usepackage{makecell}
\usepackage{hyperref}
\usepackage[capitalize]{cleveref}
\hypersetup{
    colorlinks,
    citecolor=blue,    filecolor=blue,
    linkcolor=blue,    urlcolor=blue
}

\begin{document}

\title{Measuring spatio-temporal couplings using modal spatio-spectral wavefront retrieval}

\author{N. Weiße}

\affiliation{Ludwig--Maximilians--Universit{\"a}t M{\"u}nchen, Am Coulombwall 1, 85748 Garching, Germany}%
\affiliation{Department of Physics, Clarendon Laboratory, University of Oxford, Parks Road, Oxford OX1 3PU, United Kingdom}%

\author{J. Esslinger}
\affiliation{Ludwig--Maximilians--Universit{\"a}t M{\"u}nchen, Am Coulombwall 1, 85748 Garching, Germany}%
\affiliation{Department of Physics, Clarendon Laboratory, University of Oxford, Parks Road, Oxford OX1 3PU, United Kingdom}%

\author{S. Howard}
\affiliation{Ludwig--Maximilians--Universit{\"a}t M{\"u}nchen, Am Coulombwall 1, 85748 Garching, Germany}%
\affiliation{Department of Physics, Clarendon Laboratory, University of Oxford, Parks Road, Oxford OX1 3PU, United Kingdom}%

\author{F.M. Foerster}
\affiliation{Ludwig--Maximilians--Universit{\"a}t M{\"u}nchen, Am Coulombwall 1, 85748 Garching, Germany}%

\author{F. Haberstroh}
\affiliation{Ludwig--Maximilians--Universit{\"a}t M{\"u}nchen, Am Coulombwall 1, 85748 Garching, Germany}%

\author{L. Doyle}
\affiliation{Ludwig--Maximilians--Universit{\"a}t M{\"u}nchen, Am Coulombwall 1, 85748 Garching, Germany}%

\author{P. Norreys}
\affiliation{Department of Physics, Clarendon Laboratory, University of Oxford, Parks Road, Oxford OX1 3PU, United Kingdom}%

\author{J. Schreiber}
\affiliation{Ludwig--Maximilians--Universit{\"a}t M{\"u}nchen, Am Coulombwall 1, 85748 Garching, Germany}%

\author{S. Karsch}
\affiliation{Ludwig--Maximilians--Universit{\"a}t M{\"u}nchen, Am Coulombwall 1, 85748 Garching, Germany}%

\author{A. Döpp}
\email{a.doepp@lmu.de}
\affiliation{Ludwig--Maximilians--Universit{\"a}t M{\"u}nchen, Am Coulombwall 1, 85748 Garching, Germany}%
\affiliation{Department of Physics, Clarendon Laboratory, University of Oxford, Parks Road, Oxford OX1 3PU, United Kingdom}%

\email{\authormark{*} These authors contributed equally.} 
\email{\authormark{**} a.doepp@lmu.de}

\begin{abstract}

Knowledge of spatio-temporal couplings such as pulse-front tilt or curvature is important to determine the focused intensity of high-power lasers. Common techniques to diagnose these couplings are either qualitative or require hundreds of measurements. Here we present both a new algorithm for retrieving spatio-temporal couplings, as well as novel experimental implementations. Our method is based on the expression of the spatio-spectral phase in terms of a Zernike-Taylor basis, allowing us to directly quantify the coefficients for common spatio-temporal couplings. We take advantage of this method to perform quantitative measurements using a simple experimental setup, consisting of different bandpass filters in front of a Shack-Hartmann wavefront sensor. This fast acquisition of laser couplings using narrowband filters, abbreviated FALCON, is easy and cheap to implement in existing facilities. To this end, we present a measurement of spatio-temporal couplings at the ATLAS-3000 petawatt laser using our technique.
\end{abstract}

\maketitle

%%%%%%%%%%%%%%%%%%%%%%%%%%  body  %%%%%%%%%%%%%%%%%%%%%%%%%%
\section{Introduction}

High-power laser systems based on chirped-pulse amplification (CPA) nowadays reach peak powers in excess of a petawatt\cite{Danson.2015,Danson.2019}. When focused down to their diffraction limit, the light pulses produced by these systems can reach intensities beyond $10^{23}\:\si{\watt\per\cm^2}$\cite{yoon2021realization}. At these extreme intensities, matter interacting with the laser field is immediately ionized, and both electrons and ions require a relativistic treatment, as they are accelerated to relativistic velocities within a single laser cycle\cite{mourou2006optics}. The final peak intensity does, however, depend on the spatio-temporal focusing of the laser. In contrast to CW lasers, ultrashort lasers intrinsically have a large bandwidth, and due to dispersion effects in amplification and propagation, different frequencies constituting the laser pulse do not necessarily focus in the same way. This effect can be described by spatio-temporal couplings (STCs)\cite{akturk2010spatio} in the far-field or equivalently by angular-spectral couplings in the near-field. The best-known couplings are pulse-front tilt or angular dispersion, originating for instance from wedged dispersing prisms, and pulse-front curvature, as produced by chromatic lenses\cite{jolly2020spatio}. A common source of STCs in CPA laser systems are compressors, where STCs can be generated due to misalignment or (thermal) grating deformation\cite{leroux2020description}.

In many experiments these phenomena are considered as higher-order effects and are ignored. Instead, the spatially-resolved intensity $I(x,y)$ and wavefront $\varphi(x,y)$ are measured, e.g. using a Shack-Hartmann sensor, and are combined with a single measurement of the spectral phase $\phi(\omega)$, which can for instance be measured using variants of frequency-resolved optical gating (FROG). This approximation $E(x,y,\omega)\approx \sqrt{I(x,y)}\cdot e^{i\varphi(x,y)}\cdot e^{i\phi(\omega)}$ thus assumes simultaneous focusing of the pulse in time and space. However, in facilities that rely on ultrashort laser systems, especially in high-power laser facilities that are built to explore the high-intensity frontier, these effects cannot be ignored. In addition, a number of recent papers have suggested that deliberate manipulation of the spatio-temporal focusing properties of laser pulses may be of interest for applications such as laser-plasma acceleration\cite{froula2018spatiotemporal,caizergues2020phase}.

\begin{figure*}[th]
        \centering
            \includegraphics[width=0.6\linewidth]{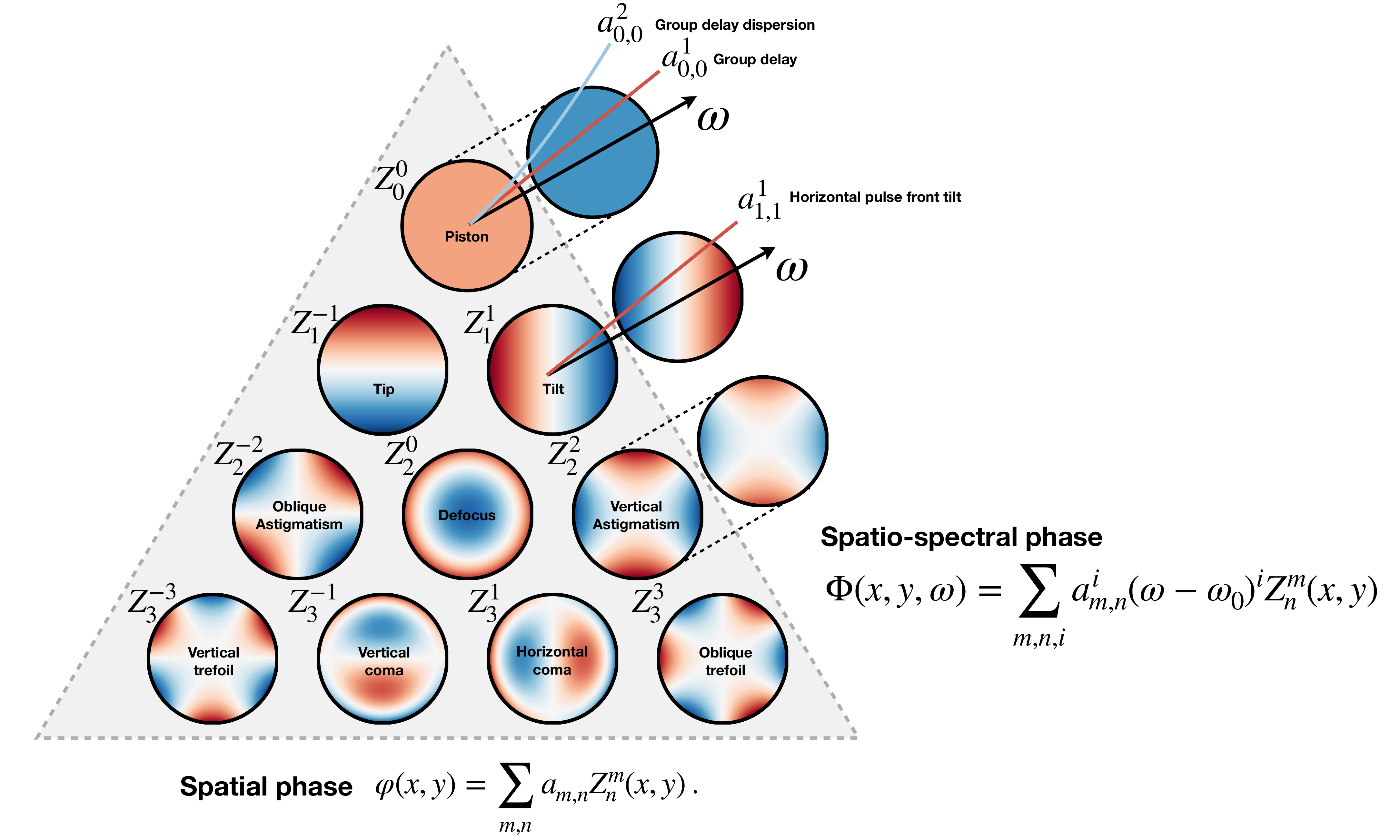}
            \caption{Visualization of the proposed basis functions to describe the spatio-spectral phase, where each Zernike mode is expanded in frequency with a Taylor series.}
            \label{fig:base_functions}
\end{figure*}

To date, techniques for measuring spatio-temporal couplings can be divided into two types, STC-specific measurements and general reconstruction techniques for the three-dimensional field. The former includes, e.g., interferometric field autocorrelation to exclusively measure pulse-front tilt (and the analogous angular chirp)\cite{pretzler2000angular} or far-field beamlet cross-correlation\cite{smartsev2022characterization} to measure pulse front tilt and curvature. The latter comprises of a variety of techniques, e.g. SEA-TADPOLE\cite{bowlan2006crossed}, STARFISH \cite{alonso_2010}, TERMITES \cite{miranda2014spatiotemporal,pariente2016space}, INSIGHT\cite{borot2018spatio}, bulk lateral shearing spectral interferometry \cite{lopez2022bulk} and so forth.
All of these techniques require some kind of spatial or temporal scanning, typically requiring hundreds or thousands of shots for a single measurement of the three-dimensional field. It should be mentioned that some attempts for single-shot measurements have been made, e.g. STRIPED-FISH\cite{gabolde2006single}, but these have some drawbacks such as a low measurement bandwidth or high experimental complexity.

In this paper we introduce a new approach for retrieving spatio-temporal couplings using modal reconstruction of the spectrally-resolved wavefront. This method requires significantly less measurement points than the aforementioned scanning techniques. Based on these requirements, we design and demonstrate FALCON, a new measurement device that is easy to operate, fast in computation, works with only a few shots and is robust in the reconstruction of the modes of interest. The paper is structured as follows: In \cref{Theory} we introduce the theory of modal laser field decomposition and describe an implementation of phase retrieval based on matrix inversion. In \cref{Experiment} we present the setup for an experimental demonstration, followed by proof-of-principle measurements with a low-power oscillator and with the ATLAS petawatt laser in \cref{sec:measurement}. \cref{sec:summary} summarizes our findings and gives an outlook on future research directions.

\section{Theory of modal reconstruction}\label{Theory}

\subsection{Laser Field Decomposition}
A general laser pulse field in a plane of observation $E(x,y,\omega)$ can be described by a spectral intensity $I(x,y,\omega)$ and a spatio-spectral phase $\Phi(x,y,\omega)$:
\begin{equation}
    E(x,y,\omega) = \sqrt{(I(x,y,\omega))}\cdot\exp{\left( i\Phi(x,y,\omega) \right)}.
\end{equation}

As already mentioned in the introduction, a common approximation is to describe the spatio-spectral phase as a combination of the spectrally-averaged wavefront $\varphi(x,y)$ and the position-independent spectral phase $\phi(\omega)$. The latter is usually described as a Taylor series,

\begin{equation}
\label{eq:spectralphasetaylorexpansion}
\phi(\omega) = \sum_{n=0}^\infty \frac{(\omega-\omega_0)^n}{n!}\cdot\left(\frac{\partial^n \phi }{\partial\omega^n}\right)_{\omega=\omega_0},
\end{equation}
where $\omega_0$ is the central frequency. The first terms of this series are known as the carrier envelope phase ($\phi_0$), group delay ($\phi_1$) and group delay dispersion ($\phi_2$). Meanwhile, the wavefront is typically described in terms of Zernike polynomials \cite{ZERNIKE1934689}
\begin{equation}
\varphi(x,y) = \sum_{m,n} a_{m,n} Z_n^m(x,y).
\end{equation}
The Zernike coefficients are defined in polar coordinates on a unit circle and the lower-order coefficients can easily be related to the descriptions introduced by Seidel \cite{seidel1857ueber}. The zeroth order, piston, is technically identical to the carrier envelope phase at the particular frequency, while first order shift terms are known as tip and tilt. Being two-dimensional, the second order terms take the forms of astigmatism, defocus and oblique (45$^{\circ}$) astigmatism.

We can expand these definitions to the general case of a spatio-spectral phase. In accordance with the definitions for both spectral phase and wavefront, we define the spatio-spectral phase $\Phi(x,y,\omega)$ as a spectral Taylor-expansion of the spatial Zernike polynomials
\begin{equation}
    \Phi(x,y,\omega) = \sum_{m,n,i}  a_{m,n}^i(\omega-\omega_0)^i Z_n^m(x,y).
    \label{eq:spatio-spectral-phase}
\end{equation}

As illustrated in \cref{fig:base_functions}, \cref{eq:spatio-spectral-phase} describes the spectral evolution of each Zernike mode. Using these definitions, the spectral phase is defined as the spectrally-resolved piston term $\phi(\omega)=\left(\sum_i a_{0,0}^i(\omega-\omega_0)^i\right) Z_0^0$ and the wavefront is given by $\varphi(x,y)=\sum_{m,n} a_{m,n}^0 Z_m^n(x,y)$. In addition, we can readily describe all common spatio-temporal couplings such as pulse-front tilt, which is determined by the $a^1_{-1,1}$ and $a^1_{1,1}$ coefficients. We would like to mention that there also exists a three-dimensional extension to the Zernike polynomials that could also be used to express $\Phi(x,y,\omega)$, but we have opted against its use in order to maintain better compatibility of our base coefficients with the common definitions of the spectral phase and spatio-temporal couplings.

\subsection{Modal retrieval algorithm}\label{algorithm}

In the following we present how the spatio-spectral basis functions defined in the previous sections can be used to reconstruct spatio-temporal couplings. Our analysis is based on the idea that many wavefront sensors (Shack-Hartmann, Shearing Interferometer, etc.) yield the local derivatives of the wavefront ($\partial\varphi/\partial x$, $\partial\varphi/\partial y$) as a measurement result.

In a \textit{zonal} reconstruction approach, single pixels that are independent of their neighbouring pixels are used as basis functions, meaning that the final wavefront is a superposition of delta functions $\Phi(x,y) = \sum_{u,v} a_{u,v}\delta(x-x_u,y-y_v)$. Here, the local gradients have to be integrated and then combined to result in the wavefront map with a resolution equal to the microlens array size. 
In contrast, the \textit{modal} approach\cite{cubalchini1979modal,southwell1980wave} reconstructs Zernike coefficients by fitting them to the measured phase gradients. This can be done by using the derivatives of the Zernike modes in cartesian coordinates, which we calculate up to the required Zernike order using the recurrence relations presented by Andersen \cite{andersen2018}. This leads not only to faster computation times, but also to more reliable results, as there are less intermediate steps required. 

While modal reconstruction is well-known in the case of wavefront analysis\cite{cubalchini1979modal,southwell1980wave}, it has to our knowledge never been applied to the more general case of reconstructing the spatio-spectral phase. Using the spatio-spectral base expansion of the phase given by \cref{eq:spatio-spectral-phase} we can easily extend the method and fit coefficients to spatio-temporal couplings such as pulse-front tilt given measurements of the phase derivatives at different frequencies. In order to implement this scheme a linear equation system is set up:
\begin{equation}
   \vec \nabla \Phi
    = \mathbf{T} \cdot \Vec{a}
\end{equation}
where $\vec \nabla \Phi$ is a vector containing the the phase derivatives $\frac{\partial \varphi}{\partial x}$ and $\frac{\partial \varphi}{\partial y}$ at all positions $(\vec x,\vec y)$ along the $x$ and $y$ axis, respectively. $\mathbf{T}$ is the forward transfer matrix that connects basis functions with the wavefront gradients and $\Vec{a}$ is a vector containing all the included coefficients of the basis functions. In the single-frequency case, on a $2\times2$ grid and including conventional Zernike coefficients up to tip and tilt, the equation system takes the form:

\begin{widetext}

\begin{equation}
    \begin{pmatrix}
        \left.\frac{\partial \varphi}{\partial x}\right|_{(0,0)} \\
        \left.\frac{\partial \varphi}{\partial x}\right|_{(0,1)} \\
        \left.\frac{\partial \varphi}{\partial x}\right|_{(1,0)} \\
        \left.\frac{\partial \varphi}{\partial x}\right|_{(1,1)} \\
        \left.\frac{\partial \varphi}{\partial y}\right|_{(0,0)} \\
        \left.\frac{\partial \varphi}{\partial y}\right|_{(0,1)} \\
        \left.\frac{\partial \varphi}{\partial y}\right|_{(1,0)} \\
        \left.\frac{\partial \varphi}{\partial y}\right|_{(1,1)}
    \end{pmatrix}
    =  
    \begin{pmatrix}
         \frac{\partial Z_0^0}{\partial x}\Big|_{0,0} &  \frac{\partial Z_1^{-1}}{\partial x}\Big|_{0,0} &  \frac{\partial Z_1^1}{\partial x}\Big|_{0,0}  \\
         \frac{\partial Z_0^0}{\partial x}\Big|_{0,1} &  \frac{\partial Z_1^{-1}}{\partial x}\Big|_{0,1} &  \frac{\partial Z_1^1}{\partial x}\Big|_{0,1}  \\
         \frac{\partial Z_0^0}{\partial x}\Big|_{1,0} &  \frac{\partial Z_1^{-1}}{\partial x}\Big|_{1,0} &  \frac{\partial Z_1^1}{\partial x}\Big|_{1,0}  \\
         \frac{\partial Z_0^0}{\partial x}\Big|_{1,1} &  \frac{\partial Z_1^{-1}}{\partial x}\Big|_{1,1} &  \frac{\partial Z_1^1}{\partial x}\Big|_{1,1}  \\
        
         \frac{\partial Z_0^0}{\partial y}\Big|_{0,0} &  \frac{\partial Z_1^{-1}}{\partial y}\Big|_{0,0} &  \frac{\partial Z_1^1}{\partial y}\Big|_{0,0} \\
         \frac{\partial Z_0^0}{\partial y}\Big|_{0,1} &  \frac{\partial Z_1^{-1}}{\partial y}\Big|_{0,1} &  \frac{\partial Z_1^1}{\partial y}\Big|_{0,1} \\
         \frac{\partial Z_0^0}{\partial y}\Big|_{1,0} &  \frac{\partial Z_1^{-1}}{\partial y}\Big|_{1,0} &  \frac{\partial Z_1^1}{\partial y}\Big|_{1,0}\\
         \frac{\partial Z_0^0}{\partial y}\Big|_{1,1} &  \frac{\partial Z_1^{-1}}{\partial y}\Big|_{1,1} &  \frac{\partial Z_1^1}{\partial y}\Big|_{1,1}
\end{pmatrix}
    \cdot
    \begin{pmatrix}
    a_{0,0} \\
    a_{1,-1} \\
    a_{1,1} 
    \end{pmatrix}
    \label{eq:T-matrix}
\end{equation}

\end{widetext}

This equation has to be solved for $\Vec{a}$. As $\mathbf{T}$ is non-square, this can be done by finding the \textit{pseudo-}inverse, $\mathbf{T^{+}}$, which is at the same time the optimal solution to the least squares formulation of the problem, i.e. $\mbox{min}\{||\mathbf{T^{+}}(\mathbf{{T}}\cdot \vec a - \vec \nabla \Phi)||^2\}$.

To also retrieve spatio-temporal coupling parameters, the forward matrix $\mathbf{T}$ must be extended by adding columns according to the desired amount of spatio-spectral parameters and by adding rows for each individual frequency-selective measurement. To give an explicit example, extending \cref{eq:T-matrix} to include the corresponding first-order spatio-spectral coefficients using measurements with two different spectral response functions $f_1(\omega)$ and $f_2(\omega)$, respectively, results in the following equation system:
\begin{widetext}
\begin{equation}
    \begin{pmatrix}
        \left.\frac{\partial \Phi}{\partial x}\right|_{(0,0,\omega_1)} \\
        \left.\frac{\partial \Phi}{\partial x}\right|_{(0,1,\omega_1)} \\
        \left.\frac{\partial \Phi}{\partial x}\right|_{(1,0,\omega_1)} \\
        \left.\frac{\partial \Phi}{\partial x}\right|_{(1,1,\omega_1)} \\
        \left.\frac{\partial \Phi}{\partial y}\right|_{(0,0,\omega_1)} \\
        \left.\frac{\partial \Phi}{\partial y}\right|_{(0,1,\omega_1)} \\
        \left.\frac{\partial \Phi}{\partial y}\right|_{(1,0,\omega_1)} \\
        \left.\frac{\partial \Phi}{\partial y}\right|_{(1,1,\omega_1)} \\
        \left.\frac{\partial \Phi}{\partial x}\right|_{(0,0,\omega_2)} \\
        \left.\frac{\partial \Phi}{\partial x}\right|_{(0,1,\omega_2)} \\
        \left.\frac{\partial \Phi}{\partial x}\right|_{(1,0,\omega_2)} \\
        \left.\frac{\partial \Phi}{\partial x}\right|_{(1,1,\omega_2)} \\
        \left.\frac{\partial \Phi}{\partial y}\right|_{(0,0,\omega_2)} \\
        \left.\frac{\partial \Phi}{\partial y}\right|_{(0,1,\omega_2)} \\
        \left.\frac{\partial \Phi}{\partial y}\right|_{(1,0,\omega_2)} \\
        \left.\frac{\partial \Phi}{\partial y}\right|_{(1,1,\omega_2)}
    \end{pmatrix}
    =  
    \begin{pmatrix}
         \frac{\partial Z_0^{0}}{\partial x}\Big|_{0,0} &  \frac{\partial Z_1^{-1}}{\partial x}\Big|_{0,0} &  \frac{\partial Z_1^1}{\partial x}\Big|_{0,0} &  \frac{\partial Z_0^0}{\partial x}\Big|_{0,0} \tilde \omega_1^1 &  \frac{\partial Z_1^{-1}}{\partial x}\Big|_{0,0}\tilde \omega_1^1 &  \frac{\partial Z_1^1}{\partial x}\Big|_{0,0} \tilde \omega_1^1\\
         \frac{\partial Z_0^{0}}{\partial x}\Big|_{0,1} &  \frac{\partial Z_1^{-1}}{\partial x}\Big|_{0,1} &  \frac{\partial Z_1^1}{\partial x}\Big|_{0,1} &  \frac{\partial Z_0^0}{\partial x}\Big|_{0,1} \tilde \omega_1^1 &  \frac{\partial Z_1^{-1}}{\partial x}\Big|_{0,1}\tilde \omega_1^1 &  \frac{\partial Z_1^1}{\partial x}\Big|_{0,1} \tilde \omega_1^1 \\
         \frac{\partial Z_0^{0}}{\partial x}\Big|_{1,0} &  \frac{\partial Z_1^{-1}}{\partial x}\Big|_{1,0} &  \frac{\partial Z_1^1}{\partial x}\Big|_{1,0} &  \frac{\partial Z_0^0}{\partial x}\Big|_{1,0} \tilde \omega_1^1 &  \frac{\partial Z_1^{-1}}{\partial x}\Big|_{1,0}\tilde \omega_1^1 &  \frac{\partial Z_1^1}{\partial x}\Big|_{1,0} \tilde \omega_1^1 \\
         \frac{\partial Z_0^{0}}{\partial x}\Big|_{1,1} &  \frac{\partial Z_1^{-1}}{\partial x}\Big|_{1,1} &  \frac{\partial Z_1^1}{\partial x}\Big|_{1,1} &  \frac{\partial Z_0^0}{\partial x}\Big|_{1,1} \tilde \omega_1^1 &  \frac{\partial Z_1^{-1}}{\partial x}\Big|_{1,1}\tilde \omega_1^1 &  \frac{\partial Z_1^1}{\partial x}\Big|_{1,1} \tilde \omega_1^1 \\
        
         \frac{\partial Z_0^{0}}{\partial y}\Big|_{0,0} &  \frac{\partial Z_1^{-1}}{\partial y}\Big|_{0,0} &  \frac{\partial Z_1^1}{\partial y}\Big|_{0,0} &  \frac{\partial Z_0^{0}}{\partial y}\Big|_{0,0}\tilde \omega_1^1 &  \frac{\partial Z_1^{-1}}{\partial y}\Big|_{0,0}\tilde \omega_1^1 &  \frac{\partial Z_1^1}{\partial y}\Big|_{0,0}\tilde \omega_1^1 \\
         \frac{\partial Z_0^{0}}{\partial y}\Big|_{0,1} &  \frac{\partial Z_1^{-1}}{\partial y}\Big|_{0,1} &  \frac{\partial Z_1^1}{\partial y}\Big|_{0,1} &  \frac{\partial Z_0^{0}}{\partial y}\Big|_{0,1}\tilde \omega_1^1 &  \frac{\partial Z_1^{-1}}{\partial y}\Big|_{0,1}\tilde \omega_1^1 &  \frac{\partial Z_1^1}{\partial y}\Big|_{0,1}\tilde \omega_1^1 \\
         \frac{\partial Z_0^{0}}{\partial y}\Big|_{1,0} &  \frac{\partial Z_1^{-1}}{\partial y}\Big|_{1,0} &  \frac{\partial Z_1^1}{\partial y}\Big|_{1,0} &   \frac{\partial Z_0^{0}}{\partial y}\Big|_{1,0}\tilde \omega_1^1 &  \frac{\partial Z_1^{-1}}{\partial y}\Big|_{1,0}\tilde \omega_1^1 &  \frac{\partial Z_1^1}{\partial y}\Big|_{1,0}\tilde \omega_1^1 \\
         \frac{\partial Z_0^{0}}{\partial y}\Big|_{1,1} &  \frac{\partial Z_1^{-1}}{\partial y}\Big|_{1,1} &  \frac{\partial Z_1^1}{\partial y}\Big|_{1,1} &  \frac{\partial Z_0^{0}}{\partial y}\Big|_{1,1}\tilde \omega_1^1 &  \frac{\partial Z_1^{-1}}{\partial y}\Big|_{1,1}\tilde \omega_1^1 &  \frac{\partial Z_1^1}{\partial y}\Big|_{1,1}\tilde \omega_1^1 \\
        
         \frac{\partial Z_0^{0}}{\partial x}\Big|_{0,0} &  \frac{\partial Z_1^{-1}}{\partial x}\Big|_{0,0} &  \frac{\partial Z_1^1}{\partial x}\Big|_{0,0} &  \frac{\partial Z_0^0}{\partial x}\Big|_{0,0} \tilde \omega_2^1 &  \frac{\partial Z_1^{-1}}{\partial x}\Big|_{0,0}\tilde \omega_2^1 &  \frac{\partial Z_1^1}{\partial x}\Big|_{0,0} \tilde \omega_2^1\\
         \frac{\partial Z_0^{0}}{\partial x}\Big|_{0,1} &  \frac{\partial Z_1^{-1}}{\partial x}\Big|_{0,1} &  \frac{\partial Z_1^1}{\partial x}\Big|_{0,1} &  \frac{\partial Z_0^0}{\partial x}\Big|_{0,1} \tilde \omega_2^1 &  \frac{\partial Z_1^{-1}}{\partial x}\Big|_{0,1}\tilde \omega_2^1 &  \frac{\partial Z_1^1}{\partial x}\Big|_{0,1} \tilde \omega_2^1 \\
         \frac{\partial Z_0^{0}}{\partial x}\Big|_{1,0} &  \frac{\partial Z_1^{-1}}{\partial x}\Big|_{1,0} &  \frac{\partial Z_1^1}{\partial x}\Big|_{1,0} &  \frac{\partial Z_0^0}{\partial x}\Big|_{1,0} \tilde \omega_2^1 &  \frac{\partial Z_1^{-1}}{\partial x}\Big|_{1,0}\tilde \omega_2^1 &  \frac{\partial Z_1^1}{\partial x}\Big|_{1,0} \tilde \omega_2^1 \\
         \frac{\partial Z_0^{0}}{\partial x}\Big|_{1,1} &  \frac{\partial Z_1^{-1}}{\partial x}\Big|_{1,1} &  \frac{\partial Z_1^1}{\partial x}\Big|_{1,1} &  \frac{\partial Z_0^0}{\partial x}\Big|_{1,1} \tilde \omega_2^1 &  \frac{\partial Z_1^{-1}}{\partial x}\Big|_{1,1}\tilde \omega_2^1 &  \frac{\partial Z_1^1}{\partial x}\Big|_{1,1} \tilde \omega_2^1 \\
        
         \frac{\partial Z_0^{0}}{\partial y}\Big|_{0,0} &  \frac{\partial Z_1^{-1}}{\partial y}\Big|_{0,0} &  \frac{\partial Z_1^1}{\partial y}\Big|_{0,0} &  \frac{\partial Z_0^{0}}{\partial y}\Big|_{0,0}\tilde \omega_2^1 &  \frac{\partial Z_1^{-1}}{\partial y}\Big|_{0,0}\tilde \omega_2^1 &  \frac{\partial Z_1^1}{\partial y}\Big|_{0,0}\tilde \omega_2^1 \\
         \frac{\partial Z_0^{0}}{\partial y}\Big|_{0,1} &  \frac{\partial Z_1^{-1}}{\partial y}\Big|_{0,1} &  \frac{\partial Z_1^1}{\partial y}\Big|_{0,1} &  \frac{\partial Z_0^{0}}{\partial y}\Big|_{0,1}\tilde \omega_2^1 &  \frac{\partial Z_1^{-1}}{\partial y}\Big|_{0,1}\tilde \omega_2^1 &  \frac{\partial Z_1^1}{\partial y}\Big|_{0,1}\tilde \omega_2^1 \\
         \frac{\partial Z_0^{0}}{\partial y}\Big|_{1,0} &  \frac{\partial Z_1^{-1}}{\partial y}\Big|_{1,0} &  \frac{\partial Z_1^1}{\partial y}\Big|_{1,0} &  \frac{\partial Z_0^{0}}{\partial y}\Big|_{1,0}\tilde \omega_2^1 &  \frac{\partial Z_1^{-1}}{\partial y}\Big|_{1,0}\tilde \omega_2^1 &  \frac{\partial Z_1^1}{\partial y}\Big|_{1,0}\tilde \omega_2^1 \\
         \frac{\partial Z_0^{0}}{\partial y}\Big|_{1,1} &  \frac{\partial Z_1^{-1}}{\partial y}\Big|_{1,1} &  \frac{\partial Z_1^1}{\partial y}\Big|_{1,1} &  \frac{\partial Z_0^{0}}{\partial y}\Big|_{1,1}\tilde \omega_2^1 &  \frac{\partial Z_1^{-1}}{\partial y}\Big|_{1,1}\tilde \omega_2^1 &  \frac{\partial Z_1^1}{\partial y}\Big|_{1,1}\tilde \omega_2^1
    \end{pmatrix}
    \cdot
    \begin{pmatrix}
    a_{0,0}^0 \\
    a_{1,-1}^0 \\
    a_{1,1}^0 \\
    a_{0,0}^1 \\
    a_{1,-1}^1 \\
    a_{1,1}^1 
    \end{pmatrix}
\end{equation}
\end{widetext}

where $\tilde \omega_k^i = \sum_\omega(\omega-\omega_0)^i f_k(\omega)$ describes the average of the spectrum seen by each lenslet for the $i$-th spatio temporal coupling order and $a_{m,n}^i$ is the coefficient of the $i$-th order Taylor expansion of the Zernike mode $a_{m,n}$, as defined in \cref{eq:spectralphasetaylorexpansion} and \cref{eq:spatio-spectral-phase}. The function $f_k(\omega)$ is therefore the normalised transmission function of the $k$-th measurement. Note that terms including $Z_0^0$ are shown in both examples for completeness, but as these terms correspond to the spectral phase $\phi(\omega)$, they have to be measured separately. 

The general $\mathbf{T}$ matrix will have $2\times n_{xy} \times n_f\times n_a$ entries, with $n_{xy}$ being the number of measurement points, $n_f$ the number of frequency data points and $n_a$ being the number of coefficients used in the reconstruction. Restricting the reconstruction to only a selected amount of Zernike and spatio-temporal coupling orders reduces the number of free parameters significantly and makes this a highly over-determined inverse problem. Not only does this significantly reduce the sensitivity to noise, but as we will see later, it also allows us to retrieve spatio-temporal coupling coefficients using only a few frequency-selective measurements. Assuming that the measurement noise acting on the individual phase gradients is normally distributed, we can estimate that the retrieval accuracy improves with $\sim {1}/{\sqrt{N}}$, with $N$ being the ratio of measurement points $n_\textrm{xy}$ to free parameters $n_\textrm{a}$. Furthermore, using the forward transfer matrix $\mathbf{T}$ we can validate the retrieval accuracy and compare the reconstructed wavefront gradients for a given frequency range with the measured ones, e.g. using the mean squared error (MSE) or median absolute deviation (MAD) as figures of merit. 

\section{Experimental setup}\label{Experiment}

As discussed in the previous section, we can use the modal reconstruction approach to directly determine spatio-temporal couplings in a laser, based on spectrally-resolved measurements of the two-dimensional wavefront gradient. There exist various experimental methods to obtain such a measurement and in the following we will focus on the combination of Shack-Hartmann wavefront sensors with both imaging Fourier transform spectroscopy and bandpass filter sets. The use of filters is particularly convenient because of its speed and simplicity. Keeping with a longstanding tradition, we name this method FALCON, as an acronym for \textit{fast acquisition of laser couplings using narrowband filters}.\footnote{We note that our experimental setup has some resemblance to the HAMSTER technique by Cousin \emph{et al.} \cite{cousin2012three}, sharing the Shack-Hartmann sensor but using an acousto-optic filter for spectral filtering and, most importantly, a different reconstruction algorithm.}

\subsection{Shack-Hartmann sensor}
\label{sec:Wavefront_sensor}
For our proof-of-principle measurement, we use home-built Shack-Hartmann (SH) wavefront sensors\cite{platt2001history}. The sensor consists of a microlens array that focuses parts of an incident laser beam as \textit{beamlets} onto a camera chip. The resulting image is a grid-like spot pattern. Using the assumption that the high-order aberrations across one lenslet are negligible, the phase is entirely dominated by the local pointing (tip and tilt) contributions. In this case the local wavefront gradient over a lenslet of pitch $d$ is given by the horizontal and vertical displacement $(\Delta x,\Delta y)$ of each spot's center of mass at a given distance $z$, typically measured in focus, multiplied by the wavenumber $k$
\begin{equation}
  \begin{pmatrix} \Delta \varphi_x / d \\ \Delta \varphi_y / d \end{pmatrix}  = \frac{k}{z}\begin{pmatrix} \Delta x \\ \Delta y \end{pmatrix}.
\end{equation}

Note that this equation uses the small-angle approximation $\tan(\theta)\simeq \theta$, which is valid since the displacement is much smaller than the focal length $f$. 

We have used two different fused silica microlens arrays (Thorlabs MLA150-5C and MLA300-14AR) with pitches of $d = \SI{150}{\micro\meter}$ and $d = \SI{300}{\micro\meter}$ and focal lengths of $f=\SI{5.6}{\mm}$ and $f=\SI{14.2}{\mm}$, respectively. The microlenses focus onto a camera sensor (Basler ace acA2040-35gm or IDS UI-5244LE-M-GL), which in our case has a pixel size of $\SI{3.45}{\micro\m}$ and $\SI{5.3}{\micro\m}$, respectively. The center of mass of each lenslet's focus can be determined with approximately 0.1 pixel accuracy, meaning that the minimum angular deviations that can be resolved are $\SI{0.345}{\micro\m}/\SI{14.2}{\mm}=\SI{24}{\micro\radian}$ with the Basler camera and the \SI{300}{\micro\meter} array and $\SI{0.53}{\micro\m}/\SI{5.6}{\mm}=\SI{95}{\micro\radian}$ with the IDS camera and \SI{150}{\micro\meter} array. The corresponding wavefront resolutions over the lenslet size are \SI{7.2}{nm} and \SI{14}{nm}, respectively, which is thus on the order of $\lambda/100$ to $\lambda/50$. Note that the angular resolutions have to be multiplied with the (de-)magnification factor in an imaging setup, where higher demagnification corresponds to higher angular resolution.

\subsection{Imaging Fourier Transform Spectroscopy}\label{IFTS}

Imaging Fourier Transform Spectroscopy (IFTS) is a common technique to obtain spatially-resolved spectral information. The method is prominently used in recent spatio-temporal characterization methods such as INSIGHT and TERMITES\cite{jeandet2020spatio}. An IFTS setup typically consists of a Michelson interferometer with one mirror placed on a motorized stage to introduce a time delay between the pulse and its replica. Both pulses are subject to spectral interference, the result of which is measured using a camera sensor, see e.g. Jeandet\cite{jeandet2020spatio} for a more detailed discussion. 

To retrieve spectral information via Fourier Transform Spectroscopy, the delay $\tau$ needs to be evenly sampled. Here we use an active-feedback piezo motor with optical encoder (Newport CONEX-SAG-LS16P). From interferometric measurements we determined that the encoder has a precision of $\sim \SI{5}{\nm}$ when moving equal steps. If not taken into account, this jitter will change the instantaneous phase, resulting in an apparent frequency jitter after the Fourier transform.

To mitigate this issue, we monitored the spectral interference during each measurement using a fiber-coupled spectrometer (Ocean Optics USB4000). We then used the Hilbert transform to extract the instantaneous phase for all wavelengths of the spectrum. Multiplied with the wavelength, this yields a precise measurement of the actual delay and we then re-sample all measurements onto a regular delay grid using third-order spline interpolation. This procedure greatly improves the measurement accuracy of the local spectrum and solves one of the most pertinent problems with Michelson-based IFTS. To ensure stability of the Hilbert transform the step size should be a fraction of the wavelength and we typically use steps of $\Delta x = \SI{25}{\nm}$, corresponding to an optical delay of $\Delta \tau = 2\Delta x/c \simeq \SI{0.167}{\fs}$. 

With a total scan range of $\SI{80}{\micro\meter}$ we get a delay range of $T = \SI{534}{\fs}$, which means that 3200 measurements are necessary, yielding a measurement time of almost an hour at 1 Hz repetition rate. The frequency resolution in IFTS is given as half the inverse of the delay range $\Delta f = 1/2T$, whereas the frequency range $F$ is determined by the step size as $F=1/2\Delta \tau$. For the parameters stated above, this evaluates to $\Delta f =\SI{ 0.94}{\tera\hertz}$ and $F = \SI{3}{\peta\hertz}$, with an equivalent wavelength resolution of about $\Delta \lambda = \SI{2}{\nm}$ around $\lambda_0 = \SI{800}{\nm}$.

\begin{figure}[t]
        \centering
            \includegraphics[width=.35\linewidth]{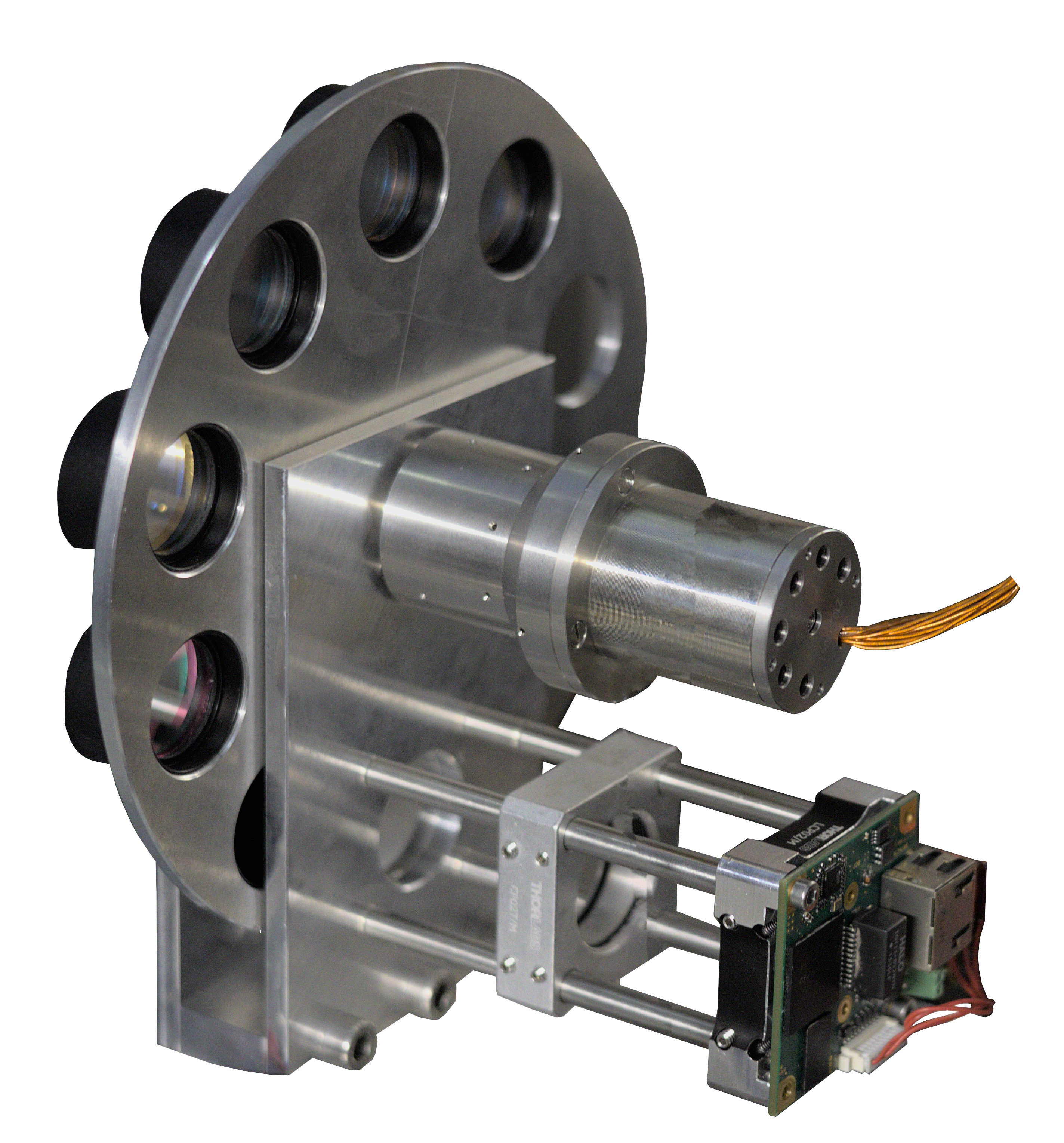}
            \includegraphics[width=.55\linewidth]{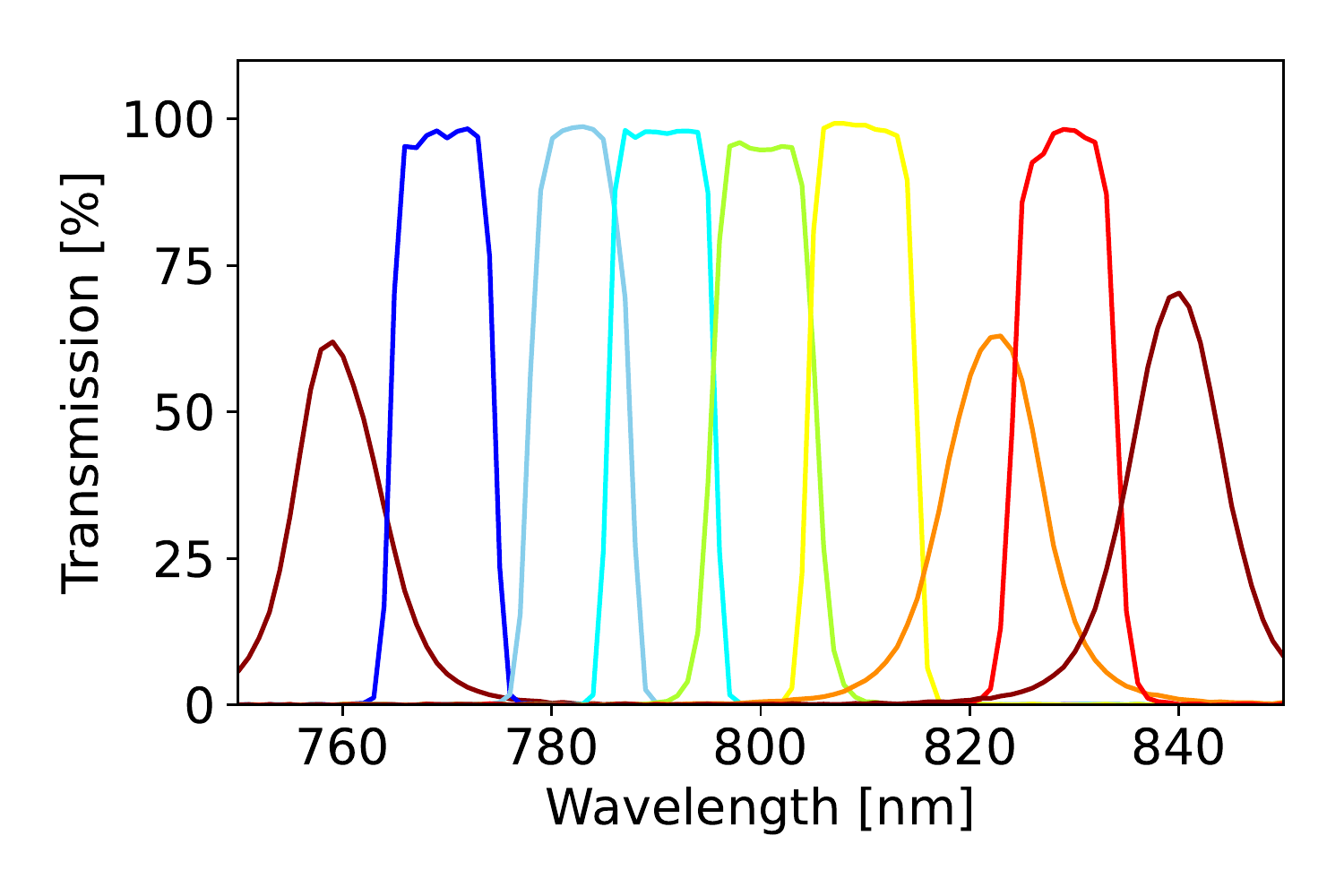}
            \caption{Left: Photograph of the experimental setup consisting of bandpass filters housed in a custom motorized wheel and a home-built Shack-Hartmann sensor. Right: Measured transmission curves of nine bandpass filters using a spectrophotometer.}
            \label{fig:Bandpass_Filter}
\end{figure}

\subsection{Bandpass filters}
\label{chap:bpfw}
The IFTS measurement described above provides high-fidelity information at the cost of a large number of acquisitions. However, to determine the dominant low-order STCs only, it is sufficient to measure the wavefront only at a few discrete frequencies. For example, two frequencies would, in principle, suffice for estimating the pulse front tilt.

To test this simplified approach, nine bandpass filters covering the spectral range from \SI{760}{nm} to \SI{840}{nm} were acquired. We used two different types of bandpass filters, each with FWHM bandwidth of \SI{10}{\nm}. The exact transmission curves $f_n(\omega)$ were measured using a spectrophotometer, see \cref{fig:Bandpass_Filter}. The selected filters cover the entire spectral range of the ATLAS-3000 laser, which we characterize with the diagnostic in \cref{sec:exp-ATLAS}. The bandpass filters are housed in a custom designed filter wheel, driven by a stepper motor, which can hold up to ten separate filters. One slot was left empty to measure the transmission without spectral filtering.

While the setup is conceptually simple to realize, in an actual implementation one has to take specific care regarding its calibration. As in any Shack-Hartmann setup, the retrieval of the wavefront can only be done relative to a reference measurement. This reference is acquired by focusing a broadband light source onto a pinhole with a diameter smaller than the focal spot size. The spatially-filtered beam transmitted through the pinhole then has a known spherical wavefront. By placing an achromatic lens at focal length distance, we can collimate this beam and we define the resulting wavefront as our reference. We take such reference images for all used filter settings to account for the transmission error of each filter.

It was found that the attenuating filters and individual spectral filters used in our setup imprint an individual, non-negligible disturbance onto the beam's wavefront. This disturbance is also not homogeneous over the entire surface of one filter, but can vary significantly when the beam is transmitted through another position of the filter. Because of this, the wavefront calibration has to be done for every filter used in the setup. Furthermore, it has to be considered that the filters are not fixed in front of the sensor, but are mounted on a rotating filter wheel. Thus, one needs to ensure that the filters can be reproducibly moved to the same positions as during calibration, which in our case is ensured by using a precise stepper motor with an angular resolution of \ang{;0.25;}. 

\subsection{Frequency-resolved Shack-Hartmann sensor}

The combination of a Shack-Hartmann sensor with either technique described in Sections \ref{IFTS} and \ref{chap:bpfw} yields frequency-resolved Shack-Hartmann images that form an array with the dimensions $(x,y,\omega)$. In IFTS this is a two-step process, as we first take measurements with different delays in the Michelson interferometer and then use a Fourier transform to obtain frequency-resolved sensor images. Bandpass filters are simpler, as they directly yield spectrally-filtered Shack-Hartmann images. In both cases, frequency-filtered slices from the three-dimensional measurement array are evaluated as described in \ref{sec:Wavefront_sensor} and the obtained gradients are then used to determine the frequency-resolved Zernike coefficients using the retrieval algorithm as described in \ref{algorithm}.

\section{Measurement results}\label{sec:measurement}

\subsection{Angular chirp introduced by wedges}
\begin{figure}[t]
    \centering
    \includegraphics[width=.9\linewidth]{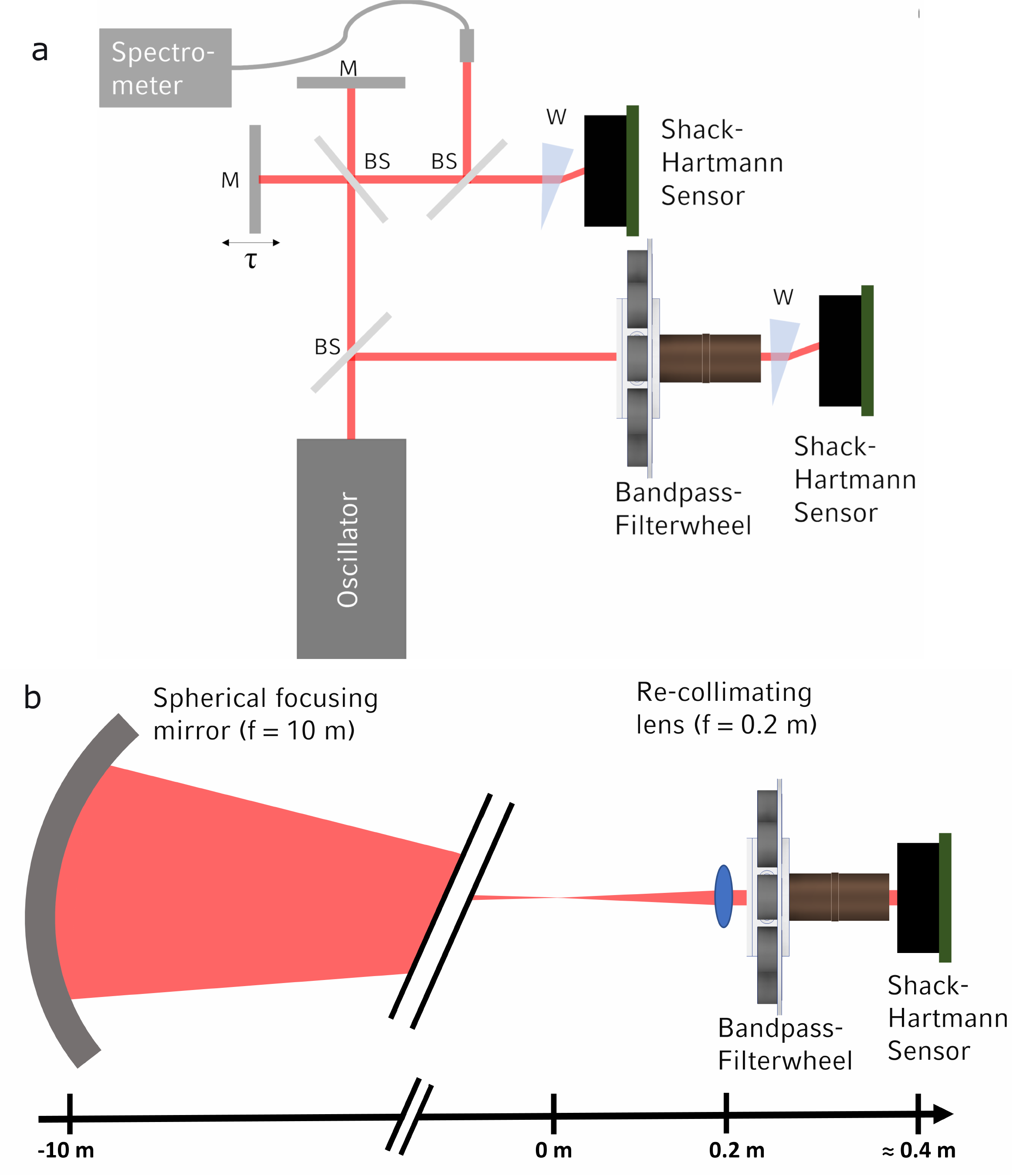}

    \caption{Experimental setup for spatio-temporal characterization using IFTS and a Shack-Hartmann sensor shown for the wegde measurement in (a) and for the ATLAS 3000 characterization in (b). The angular chirp introduced by a wedge (W) is characterised by the FALCON setup and IFTS. The additional optical components are labeled M for mirror and BS for beamsplitter.}
    \label{fig:Sketch_Experimental_Setup_FTS}
\end{figure}

In order to test our retrieval method, we used a broadband fs-oscillator as input and measured the angular chirp introduced by two different BK7 wedges with apex angles of $\alpha_1 = \ang{11;22;}$ and $\alpha_2 = \ang{18;9;}$. The wedges are places in the collimated laser beam, directly in front of the Shack-Hartmann sensor (using the Thorlabs MLA300-14AR microlens array with the Basler ace acA2040-35gm camera). The experimental setup is sketched in \cref{fig:Sketch_Experimental_Setup_FTS}, including the Michelson interferometer used for IFTS as described in \cref{IFTS}. For the setup combining IFTS with a Shack-Hartmann sensor, the result at \SI{800}{nm} was taken as a reference; for the FALCON setup separate reference images were taken. For the FALCON setup we used a subsection of the filters described in \cref{chap:bpfw}, namely the \SI{760}{nm}, \SI{790}{nm}, \SI{800}{nm}, \SI{820}{nm} filters and an additional \SI{850}{nm} filter.

\begin{figure}[tb]
    \centering
        \includegraphics[width=.9\linewidth]{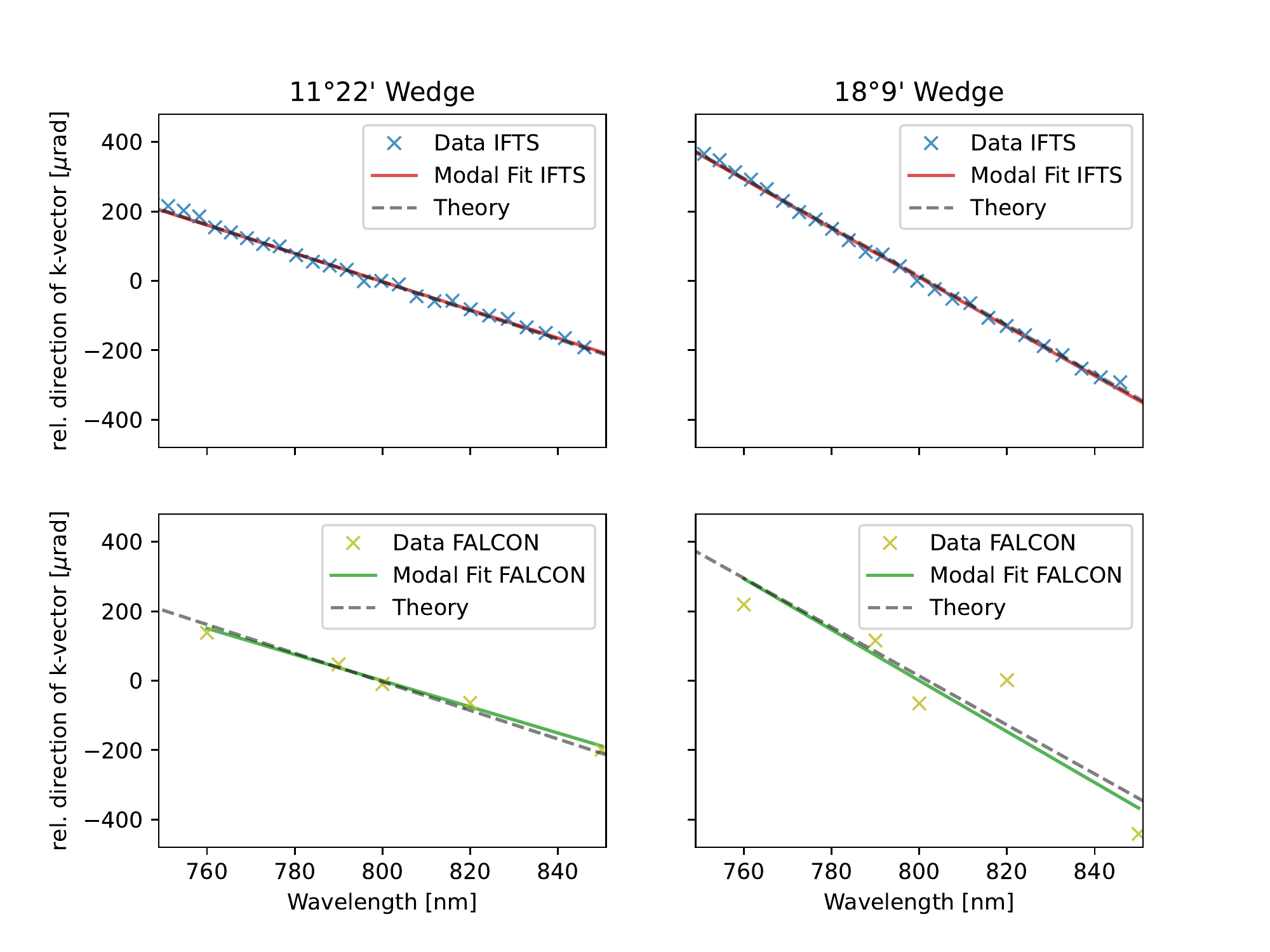}
    
        \begin{tabular}{lccc}
        \hline
        \textbf{Apex Angle} & $\alpha_1 = \ang{11;22;}$ & $\alpha_2 = \ang{18;9;}$ & \\
        \hline
        \textbf{Theoretical} [\si{\micro\radian\per\nano\metre}] & $4.12\pm 0.1$ & $7.05 \pm 0.1$& \\
        \textbf{IFTS+SH} [\si{\micro\radian\per\nano\metre}] & 4.07 & 7.08 & \\
        \textbf{FALCON} [\si{\micro\radian\per\nano\metre}] & 3.77 & 7.34 & \\
        \hline
        \end{tabular}

        \caption{Theoretical values and measured values as well as the modal fits for the angular chirp relative to the central wavelength of $\SI{800}{\nano\meter}$ as introduced by a $\ang{11;22;}$ and a $\ang{18;9;}$ wedge. Top shows the result for the IFTS measurement, bottom for the FALCON setup with only five filters used.}
        \label{fig:IFTS_SH_Results}
\end{figure}

An advantage of this setup is that the angular dispersion and hence, pulse-front tilt, introduced by a wedge can be accurately calculated using simple geometrical optics. A beam of light entering a BK7 wedge perpendicular to its first surface will be refracted at an angle $\xi(\lambda) = \arcsin\left( n(\lambda)\cdot \sin(\alpha) \right) - \alpha$. At a central wavelength of $\SI{800}{\nm}$, the theoretical value of the angular dispersion for BK7 is approximately $\SI{4.12}{\micro\radian\per\nano\metre}$ and $\SI{7.05}{\micro\radian\per\nano\metre}$ for $\alpha_1 = \ang{11;22;}$ and $\alpha_2 = \ang{18;9;}$, respectively, with an uncertainty of $\pm\SI{0.1}{\micro\radian\per\nano\metre}$ related to the alignment precision of the wedge surface.

As shown in \cref{fig:IFTS_SH_Results}, the results from the IFTS and the FALCON measurement agree very well with these theoretical estimates. The angular chirp using the direct modal retrieval and the IFTS is $\SI{4.07}{\micro\radian\per\nano\metre}$ and $\SI{7.08}{\micro\radian\per\nano\metre}$, respectively, which is in agreement with the theoretical prediction and within the estimated uncertainty. Using the FALCON measurement the reconstructed values are $\SI{3.77}{\micro\radian\per\nano\metre}$ and $\SI{7.34}{\micro\radian\per\nano\metre}$, respectively. We attribute the discrepancy between the two measurement to the fact that the FALCON was not self-referenced. Instead, we measured the wavefront with and without the wedge, keeping the Shack-Hartmann sensor in the same position. As the wedge shifts the beam only a smaller part of the signal could be evaluated, increasing the influence of noise in this particular configuration. Additionally, the IFTS took 3200 measurements, the FALCON only one measurement for each of the five filters. We thus expect that the IFTS has an approximately 25 times smaller measurement uncertainty.

\subsection{Characterization of STCs in a petawatt-class laser}\label{sec:exp-ATLAS}

Having verified our method in the previous section, we are now presenting a characterization of spatio-temporal couplings of the ATLAS-3000 laser at the Centre of Advanced Laser Application (CALA) in Garching, Germany, using the FALCON setup described in \cref{chap:bpfw}.

The ATLAS-3000 laser is a high-peak-power Ti:Sa laser system delivering a maximum pulse energy of 90 Joules before compression, at a repetition rate of \SI{1}{\hertz}. The laser pulse can either be delivered with full energy for particle acceleration experiments, or for diagnostic purposes, with greatly reduced energy by using an attenuator placed behind the final amplifiers and before the laser beam expander and compressor. For the measurements presented in the following, the laser was operated at 19.5 Joules before compression. In order to implement our diagnostic we had to ensure that the attenuated laser beam still contains the entire laser spectrum. In the past, the attenuated beam was the leakage through two dielectric mirrors coated for the laser wavelength, which predominantly transmitted the edges of the laser spectrum. The new attenuator uses the reflection of two uncoated wedges and thus mostly conserves the laser spectrum. 

Measurements are performed on a de-magnified image of the laser's near field. For this purpose we use a telescope consisting of a spherical mirror with $f_1=\SI{10}{\metre}$ focal length, otherwise used for electron acceleration experiments, combined with an $f_2=\SI{0.2}{\metre}$ achromatic doublet lens. The laser enters the spherical mirror under a small angle and the resulting aberrations are pre-corrected with a deformable mirror. The telescope images a plane of the near field approximately \SI{5}{\metre} before the spherical mirror. The intermediate focus of the telescope has a size of approximately \SI{50}{\micro\metre} diameter. To calibrate the phase measurement we place a motorised pinhole with \SI{20}{\micro\metre} diameter at the focus. The beam is fully imaged on the in-vacuum Shack-Hartmann sensor with a beam diameter of \SI{4}{mm}, corresponding to a de-magnification of $1:50$. Using the Thorlabs MLA150-5C microlens array and the IDS UI-5244LE camera with a pixel size of $\SI{5.3}{\micro\meter}$, we estimate an angular sensitivity of the order of \SI{2}{\micro\radian} in this measurement configuration. 

\begin{figure}[t]
    \centering
        \includegraphics[width=.9\linewidth]{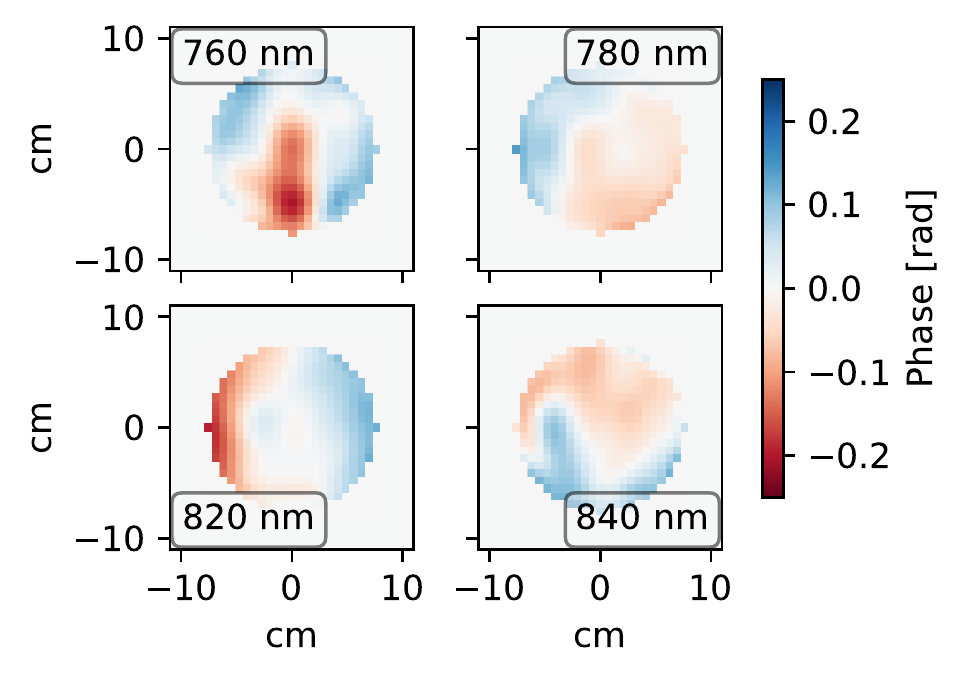}
            \begin{tabular}[t]{ l  c}
        \hline
         \textbf{Pulse Front Tilt} [$\mu \mathrm{rad}/\mathrm{nm}$] & 0.023 \\
         \hline
         \textbf{Pulse Front Curvature}
 [$\mathrm{mrad}^2/\mathrm{nm}$] & -0.0068 \\
         \hline
    \end{tabular}

        \caption{Top: Retrieved spectrally-resolved phase maps of the ATLAS-3000 laser for four different wavelengths as difference to the wavefront at the central wavelength of \SI{800}{nm}. Bottom: Retrieved values for the pulse front tilt and curvature parameters.}
        \label{fig:Atlas_stc_result}
\end{figure}

For each filter setting, five wavefront images are taken at 1 Hz repetition rate and averaged. The acquisition time for an entire measurement is thus of the order of  a minute. The resulting Shack-Hartmann images can be evaluated in two different ways. One can either fit only the coefficients of interest, e.g. $a_{-1,1}^1$ and $a_{1,1}^1$ of the vertical and horizontal angular chirp, respectively. Or, in order to describe the spectral-phase as accurately as possible, by fitting many coefficients until the validation error reaches the resolution limit. We use the median absolute deviation (MAD) as a figure of merit, as it yields the same units as the resolution and is less sensitive to noise outliers than the MSE.

For the simple case, the frequency-independent term of the wavefront was fitted up to the 24th order of Zernike polynomials, whereas only the coupling terms coefficients for the pulse front tilt and the pulse front curvature were added. The validation error for this case is $\sigma_{MAD}=\SI{0.130}{pixels}$, which is of the same order as our estimated resolution limit. The values for both the horizontal and vertical angular chirp are \SI{0.024}{\micro\radian\per\nano\meter}, which is close to the resolution limit of the diagnostic. Over a bandwith of \SI{50}{\nano\meter} and a focal length of \SI{10}{\meter} this corresponds to a shift of the focus of about \SI{12}{\micro\meter}, which is less than the typical focal spot size of $\sim \SI{50}{\micro\meter}$ in an $f/50$ configuration. The pulse front curvature is \SI{-0.0078}{\milli\radian^2\per\nano\meter}.

Fitting up to 24th order Zernike and the spatio temporal couplings of 10 orders of Zernikes up to the $\omega^5$ term 
further lowers the validation error to $\sigma_{MAD}=\SI{0.116}{pixels}$. This indicates a small, but non-negligible effect of higher-order coupling terms. \cref{fig:Atlas_stc_result} shows how the wavefront differs to the wavefront at \SI{800}{\nano\meter}. Spatio-temporal couplings are below $\lambda/10$ in the wavelength region between 780 and \SI{820}{\nano\meter} - where the spectral intensity of the laser is the largest - and hence, are well compensated. 

\section{Summary and outlook}\label{sec:summary}

We have presented a new retrieval method for spatio-temporal couplings using a modal reconstruction approach. Due to the denser forward matrix, our method is more robust to noise than established zonal reconstruction methods and allows us to retrieve spatio-temporal coupling coefficients using only a small number of frequency-selective measurements. The filter-based FALCON method is simple to implement and wavefront sensors in existing facilities can easily be upgraded at low cost. The device was implemented at the ATLAS-3000 petawatt laser to characterize spatio-temporal couplings. Furthermore, we have introduced and benchmarked an implementation based on imaging Fourier transform spectroscopy, which allows for precision measurements, albeit at the prize of significantly longer measurement times. 

While our discussion has concentrated on the retrieval of spatio-temporal couplings of the phase in the nearfield, our device is equally capable of measuring amplitude couplings. Combined with a single measurement of the spectral phase it can be easily used to retrieve the spatio-temporal electric field. Future developments may include a further increase in wavefront resolution, adaptive coefficient selection in the fitting process based on sparsity constraints \cite{dopp2022data}, and the development of a single-shot version of the method to explore shot-to-shot fluctuations of STCs.\\

\textit{Acknowledgments.} This work was supported by the Independent Junior Research Group "Characterization and control of high-intensity laser pulses for particle acceleration", DFG Project No.~453619281. N.W. was supported via the IMPULSE project by the European Union Framework Program for Research and Innovation Horizon 2020 under grant agreement No.~871161. F.M.F.\ is part of the Max Planck School of Photonics supported by BMBF, Max Planck Society, and Fraunhofer Society. P.N.\ acknowledges support via the UKRI-STFC grant ST/V001655/1. \\

\end{document}